\documentclass[aps,pra,twocolumn]{revtex4}
\usepackage{amssymb,amsthm,amsmath,enumerate,url,graphicx}
\theoremstyle{plain}
\newtheorem{theorem}{Theorem}

\newtheorem{corollary}[theorem]{Corollary}

\theoremstyle{definition}

\def\FCW{1.0\columnwidth}

\newcommand{\M}{\mathcal{M}}
\newcommand{\R}{\mathcal{R}}
\newcommand{\Tr}{\mathrm{Tr}}
\newcommand{\Lik}{\mathcal{L}}
\renewcommand{\P}{\mathcal{P}}

\newcommand{\ket}[1]{\ensuremath{\left|\scriptstyle#1\right\rangle}}
\newcommand{\bra}[1]{\ensuremath{\left\langle\scriptstyle#1\right|}}

\newcommand{\ketbra}[2]{\ket{#1}\!\!\bra{#2}}
\newcommand{\expect}[1]{\ensuremath{\left\langle#1\right\rangle}}
\newcommand{\braopket}[3]{\ensuremath{\bra{#1}#2\ket{#3}}}
\newcommand{\proj}[1]{\ketbra{#1}{#1}}

\def\Id{1\!\mathrm{l}}

\newcommand{\rhohat}{\hat{\rho}}
\newcommand{\rhoBME}{\rhohat_{\scriptscriptstyle\mathrm{BME}}}
\newcommand{\rhoMLE}{\rhohat_{\scriptscriptstyle\mathrm{MLE}}}
\newcommand{\rhotomo}{\rhohat_{\scriptscriptstyle\mathrm{tomo}}}

\newcommand{\phat}{\hat{p}}

\newcommand{\diff}{\mathrm{d}\!}
\newcommand{\diffk}[1]{\mathrm{d}_{#1}\!}

\begin{document}
\author{Robin Blume-Kohout}
\affiliation{Institute for Quantum Information, Caltech 107-81, Pasadena, CA 91125 USA}

\title{Optimal, reliable estimation of quantum states}

\begin{abstract}
Accurately inferring the state of a quantum device from the results of measurements is a crucial task in building quantum information processing hardware.  The predominant state estimation procedure, maximum likelihood estimation (MLE), generally reports an estimate with zero eigenvalues.  These cannot be justified.  Furthermore, the MLE estimate is incompatible with error bars, so conclusions drawn from it are suspect.  I propose an alternative procedure, Bayesian mean estimation (BME).  BME never yields zero eigenvalues, its eigenvalues provide a bound on their own uncertainties, and it is the most accurate procedure possible.  I show how to implement BME numerically, and how to obtain natural error bars that are compatible with the estimate.  Finally, I briefly discuss the differences between Bayesian and frequentist estimation techniques.
\end{abstract}
\date{\today}

\maketitle

One of the prerequisites for quantum computing is ``The ability to initialize the state of the qubits to a simple fiducial state, such as $\ket{000\ldots}$'' \cite{DiVincenzoFDP00}.  The device that prepares such a state must be tested and characterized, either to confirm that it reliably produces $\ket{000\ldots}$, or to determine what state it \emph{does} produce, so that it can be tuned to emit the desired one.  This task, of experimentally finding a density matrix $\hat{\rho}$ to describe the output of a quantum device, is \emph{quantum state estimation}.

State estimation is more generally useful than it may appear.  Two of the other quantum computing building blocks listed in DiVincenzo's seminal paper \cite{DiVincenzoFDP00} (low-noise universal gates, and minimal decoherence) refer to quantum \emph{processes}.  Quantum process estimation, used to characterize gates and decoherence, is mathematically equivalent to state estimation \cite{AltepeterPRL03}.  Thanks to quantum error correction and fault tolerant design, states (e.g., of the ancillae used for error correction) and gates for quantum computing need not be perfect, nor does the designer have to characterize them with infinite precision.  They must function correctly with probability at least $1-\epsilon$, where $\epsilon$ (the fault tolerance threshold) is thought to be somewhere between $10^{-5}$ \cite{AliferisQIC06} and $10^{-2}$ \cite{KnillNature05}.  A procedure for state estimation must accurately estimate probabilities on the order of $\epsilon$, and must provide a reliable bound on the uncertainty in the estimate.

\emph{Maximum likelihood estimation} (MLE), based on the principle that \textbf{the best estimate is the state $\rho$ that maximizes the probability of the observed data}, is the current procedure of choice.  Unfortunately, it has serious flaws.  It typically yields a rank-deficient estimate, with one or more zero eigenvalues.  Such an estimate is implausible, implying that some measurement outcome is literally impossible.  No finite amount of data can justify such certainty.  More importantly, it is impossible to bracket a zero probability with consistent error bars \footnote{The savvy reader may be wondering about the error-estimation procedure proposed for MLE in \cite{HradilPRA97}, and used (e.g.) in \cite{AltepeterAAMOP05}.  These error bars are the variance of many MLE estimates, on many datasets obtained by simulating measurements on the original $\rhoMLE$.  They are simply not compatible with the [rank-deficient] estimate.  Are they \emph{good} error bars -- i.e., accurate, and compatible with some better estimate?  A full answer to this question is in preparation \cite{RBKErrorBarsInPrep}; the short answer is: \emph{no}, because this procedure computes the Fisher information matrix \emph{at $\rhoMLE$}, not the true state.  Example: we flip a coin once (getting, w/o l.o.g., ``heads''), and use MLE to estimate $\phat_{\mathrm{heads}}=1$.  All the simulated measurements yield ``heads'', leading to the absurd conclusion that $\Delta p=0$.}.  The MLE estimate is at best sub-optimal, and at worst dangerously unreliable (implying, for instance, that certain errors can be ruled out).

\emph{Bayesian mean estimation} (BME) is an alternative procedure that avoids these pitfalls.  Unlike MLE, which seeks a unique maximally plausible state, BME considers the other states that are only slightly less plausible.  The simple underlying principle is that \textbf{the best estimate is an average over all states $\rho$ consistent with the data, weighted by their likelihood}.  The BME estimate is always full-rank, and comes equipped with a natural set of error bars.  Moreover, each eigenvalue $\lambda$ of $\rhoBME$ yields an upper bound ($\Delta\lambda^2\leq\lambda$) on its own uncertainty.  Best of all, BME is provably the most accurate scheme possible \cite{RBK06}, under certain reasonable assumptions.


The body of this paper is divided into three sections.  Section \ref{secMLE} explains the problems with MLE.  Section \ref{secBME} presents and analyzes the BME algorithm, along with one possible implementation.  Section \ref{secExtensions} discusses some unsolved problems.

\section{The State of the Art} \label{secMLE}

The oldest and simplest estimation procedure is \emph{quantum state tomography}.  In tomography, the estimator repeatedly measures several observables, records the frequencies of the outcomes, and identifies the outcomes' frequencies with their probabilities.  Inverting Born's Rule yields a unique density matrix $\rhotomo$ that predicts these probabilities.  The most important problem with tomography is that $\rhotomo$ often has negative eigenvalues, which means it cannot represent a physical state.

In 1996, Hradil proposed \emph{maximum likelihood estimation} (MLE) as a more flexible and sophisticated approach \cite{HradilPRA97}.  An estimate $\hat{\rho}$ is a theory about the unknown state.  Statisticians define the \emph{likelihood} of a theory, $\Lik(\rho)$ as the probability that the theory ($\rho$) would have predicted for the observed data ($\M$), before the experiment took place:
\begin{equation}
\Lik(\rho) \equiv p(\M|\rho).
\end{equation}
Thus, $\rhoMLE$ is simply the $\rho$ that maximizes $\Lik(\cdot)$ -- i.e, the most ``likely'' state.  $\Lik(\rho)$ is \emph{not} a probability distribution over $\rho$, so $\rhoMLE$ is not in any well-defined sense the most probable state.  Actually finding $\rhoMLE$ requires numerics, but several algorithms exist \cite{HradilLNP04}.  MLE was successfully applied in 2001 to a quantum optics experiment \cite{JamesPRA01}, and has been used extensively since then.

MLE has some critical flaws.  The most visible is that $\rhoMLE$ can be rank-deficient.  If $\ket{\psi}$ is the eigenstate corresponding to a zero eigenvalue, then $\braopket{\psi}{\rho}{\psi}=0$.  Such an estimate, though perhaps not unphysical, is \emph{implausible} -- i.e., no experimentalist would believe it.  It predicts exactly zero probability for every measurement outcome $\proj{\psi}$ such that $\braopket{\psi}{\rho}{\psi}=0$.  This implies \emph{absolute} certainty that $\proj{\psi}$ will not be observed, which cannot be justified by a finite amount of data.  If $N$ observations with $d$ possible outcomes are available, then the lowest defensible probability estimate for any event is roughly \footnote{At first glance, $\phat_{\mathrm{min}} = 1/N$ might seem more natural.  However, consider rolling a 100-sided die, just twice.  Assigning $\phat\approx 1/N = \frac12$ to 98 unobserved outcomes is impossible, but $\phat\approx\frac{1}{102}$ makes sense.}
\begin{equation}
\phat_{\mathrm{min}} \approx \frac{1}{N+d}. \label{eqPMin}  
\end{equation}

This is a practical concern.  One of the seminal papers on MLE \cite{JamesPRA01} estimated the polarization state of two entangled photons, produced by parametric downconversion.  The estimated $4\times 4$ density matrix has 2 eigenvalues that are exactly zero.  More recently, MLE was used to estimate the entangled state of 8 ionic qubits in a trap \cite{HaeffnerNature05}. Of 256 eigenvalues, more than 200 are less than $1/N$ (about $10^6$ measurements were made), and at least 80 are zero (to within machine precision).

Zero eigenvalues are just the most extreme illustration of a more general problem; $\rhoMLE$ implies predictions that cannot be justified by the data.  After $100$ observations, $p=10^{-8}$ is no more credible than $p=0$.  To put it another way, taking $\rhoMLE$ seriously might be exceedingly embarrassing in light of \emph{further} data.  Viewed this way, zero eigenvalues are just a symptom of the larger problem.  However, they motivate some useful questions that lend structure to this analysis:
\begin{enumerate}
\item Why are zero eigenvalues a problem?
\item Why does MLE produce zero eigenvalues?
\item What is the underlying problem with MLE?
\end{enumerate}

\subsection{Why are zero eigenvalues a problem?} \label{secMLEImplausibility}

A quantum state is nothing more or less than a prediction of the future.  Like a classical probability distribution, it predicts probabilities for all measurements that could be performed.  A state estimate is the estimator's best prediction of what future experimentalists will find when they observe a copy of the estimated system.  We should therefore evaluate an estimate on how well it predicts the future.  

Quantitative evaluation of an estimate is subject to debate.  Statisticians disagree about how to interpret even a simple statement about a coin flip: ``The probability of observing `tails' is $p_\mathrm{tails}=\frac34$.''  However, it \emph{is} indisputable that ``$p_\mathrm{tails}=0$'' implies that ``tails'' will never be observed, and that ``$p_\mathrm{tails}=1$'' implies that nothing but ``tails'' will ever be observed.  Such a statement has the force and status of a mathematical theorem, just like ``There is no largest prime number.''

An estimator should hardly claim ``$p_\mathrm{tails}=0$'' just because she has never observed ``tails''.  She has a finite number of observations to work from, and $p_\mathrm{tails}$ might simply be very small.  For example, if the data comprise a single flip, then at least one of the possible outcomes will never have been observed, but this does not justify asserting that it will \emph{never} occur.  Even if a dozen trials all yield heads, ``$p_\mathrm{tails}=0$'' is unjustified.  No matter how many data points the estimator has, we can always imagine a much larger dataset in the future, which might [embarrassingly] debunk the prediction ``$p=0$''.  Thus, data can never justify reporting $\phat=0$. Only prior knowledge, such as an impossibility theorem, can do so.

One might object that an estimate carries with it an implied uncertainty.  For instance, $\phat=0.5$ is clearly a decent estimate of $p=0.51$; why is $\phat=0$ not an equally good estimate of $p=0.01$?
The reason is that \textbf{zero probabilities are not compatible with \emph{any} error bars} 
\footnote{In addition to the explanation given in the text, there is another reason why an error of $\Delta p = 0.01$ is acceptible for $\phat=0.5$, but not for $\phat=0$.  The canonical application of probabilities is in making bets; if event $E$ has probability $p$, then a canny bettor is justified in accepting odds of $(p^{-1}-1):1$ or better against its occurrence.  When $p=0.51$, a bettor who accepts $1:1$ odds will slowly lose his money -- but when $p=0.01$, the bettor who believes $\phat=0$ and accepts $\infty:1$ odds is truly courting disaster!  The operational penalty for believing an [even slightly] erroneous estimate of $p=0$ is, indeed, severe.}.
The estimate $\phat=0.5$ could mean $\phat=0.5\pm0.01$, meaning ``$p$ is probably between 0.49 and 0.51.''  To report $\phat=0\pm0.01$, however, is nonsensical.  This would mean ``$p$ is probably between -0.01 and 0.01,'' but because $p$ must be non-negative, an unconditionally better description is ``$p$ is probably between 0 and 0.01,'' or $\phat=0.05\pm0.05$.

This is not the only way of representing ``$p$ is probably between 0 and 0.01.''  If the estimator's confidence is skewed toward one side of the interval, then the best $\phat$ might not be at its center.  However, it \emph{should} necessarily be within the interval, not on its boundary.  An estimate on the boundary can't be optimal, because moving the estimate inside the boundary by some tiny $\epsilon$ improves it (even if the optimal $\epsilon$ is unknown).  Since $p=0$ is on the boundary of any interval, $\phat=0$ is only optimal when the confidence interval has zero width.  Taken seriously, a zero probability thus implies both: (1) absolute certainty about the outcomes of future measurements; and (2) absolute certainty about the probability itself.

This has practical consequences.  If we accept that zero probabilities are implausible, then each zero eigenvalue in $\rhohat$ should be replaced by a small, but finite, $\epsilon$.  This poses two substantial problems.  First, what is $\epsilon$?  It clearly declines with $N$, but whether it should scale as $1/N$ or $1/\sqrt{N}$ is unclear.  Moreover, when statistics from many distinct observables are collated, it's not clear what $N$ is.  Second, how does ``fixing'' $\rhohat$'s small eigenvalues affect its large eigenvalues?  Since $\Tr\rho=1$ is fixed, increasing many small eigenvalues will require decreasing the largest ones.  These large eigenvalues are critical to most of the quantities of interest -- entanglement, gate fidelity, etc.  The only way to resolve this messy situation is to avoid zero eigenvalues in the first place.

\subsection{Why does MLE produce zero eigenvalues?} \label{secMLEWorkings}

The zero eigenvalues in $\rhoMLE$ are connected to the negativity of tomographic estimates.  What I will show in this section is that, for a given dataset, \textbf{if $\rhotomo$ is not positive, then $\rhoMLE$ is rank-deficient.}  On the other hand, if the tomographic estimate \emph{is} positive, then $\rhoMLE = \rhotomo$.  MLE is thus a sort of ``corrected tomography''\footnote{This discussion is rigorously correct only when the estimator measures a complete -- rather than overcomplete -- set of observables.  For an overcomplete set, it's not entirely obvious what ``tomography'' means.  The most common answer would be a least-squares fit, in which case $\rhotomo\geq0$ does not imply $\rhotomo=\rhoMLE$.  The general conclusions still hold, however, and rigor can be retained by defining ``tomography'' to mean unconstrained MLE, for overcomplete data.}.

The valid \emph{state-set}, comprising all positive density matrices, is a convex subset of Hilbert-Schmidt space, the $(d^2-1)$-dimensional vector space of Hermitian, trace-1 matrices.  Its boundary comprises the rank-deficient states.  Whenever $\rhotomo$ lies outside this boundary, MLE squashes it down onto the boundary, producing a rank-deficient estimate.  To demonstrate the connection, we begin by reviewing tomography.

\subsubsection{How tomography works} \label{secMLEWorkingsTomo}

Quantum state tomography is based on inverting Born's Rule:  \textbf{If a POVM measurement $\P = \{E_1\ldots E_N\}$ is performed on a system in state $\rho$, then the probability of observing $E_i$ is $p_i = \Tr(E_i\rho)$.}  The probabilities for $d^2$ linearly independent outcomes single out a unique $\rhotomo$ consistent with those probabilities.  Several projective measurements (at least $d+1$) can, in aggregate, form a \emph{quorum} -- i.e., provide sufficient information to identify $\rhotomo$.

Note, however, that no measurement can reveal the \emph{probability} of an event.  Repeated measurements yield \emph{frequencies}, from which the tomographic estimator infers probabilities.  The measurement is repeated $N$ times, and if outcome $E_i$ appears $n_i$ times, we estimate $\phat_i = n_i/N$.  If the measurements form a quorum, then the equations
\begin{equation}
\Tr\left(\rhotomo E_i\right) = \frac{n_i}{N}
\end{equation}
can be solved to yield a unique $\rhotomo$.

Tomography thus seeks a density matrix whose predictions agree exactly with the observed frequencies.  Unfortunately, this matrix is not always a state.  Suppose that an experimentalist, estimating the state of a single qubit, measures $\sigma_x$, $\sigma_y$, and $\sigma_z$ -- but only one time each!  Having observed the $+1$ result in each case, she seeks a $\rhotomo$ satisfying $\expect{\sigma_x}=\expect{\sigma_y}=\expect{\sigma_z}=1$.  Such a matrix exists,
\begin{equation}
\rhotomo = \left(\begin{array}{cc} 1 & \frac{1+i}{2} \\ \frac{1-i}{2} & 0 \end{array}\right),
\end{equation}
but it has a negative eigenvalue $\lambda = \frac{1-\sqrt2}{2} \approx -0.207$.  Moreover, this ``state'' implies that all three spin measurements would be perfectly predictable, which is impossible.

\begin{figure}[tb]
\includegraphics[width=\FCW]{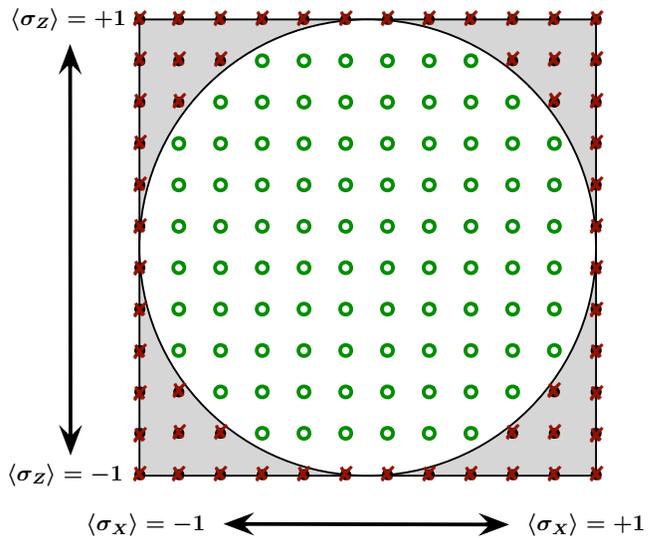}
\caption{A cross-section of the ``Bloch cube'', which contains all the possible tomographic estimates, and circumscribes the Bloch sphere containing all \emph{positive} estimates.  The points shown are possible tomographic estimates for $N=11$ measurements each of $\sigma_x$ and $\sigma_z$, with $\expect{\sigma_y}$ set to zero for simplicity's sake.  Of the 144 $\rhotomo$ shown, 54 are non-positive (keep in mind that the $\sigma_y$ dimension is ignored).  Depending on the state, some $\rhotomo$ will of course be more likely than others; this figure merely illustrates the array of \emph{possible} non-positive estimates.}
\label{figBlochCube}
\end{figure}

Estimating the state from a single measurement of each basis is a rather extreme example.  However, it illustrates a point.  Tomography, in a single-minded quest to match Born's Rule to observed frequencies, pays no attention to positivity.  As the number of measurements ($N$) increases, the possible tomographic estimates form an $N\times N\times N$ grid.  They fill a ``Bloch cube,'' defined by $\expect{\sigma_{x,y,z}}\in[-1\ldots1]$, which circumscribes the Bloch sphere and contains a lot of negative states (see Fig. \ref{figBlochCube}).  If the true state is sufficiently pure, then the probability of obtaining a negative estimate can remain as high as $50\%$ for arbitrarily large $N$, since the true state is very close to the boundary between physical and unphysical states.

In larger Hilbert spaces, the problem gets worse for two reasons.  First, the state-set's dimensionality (and therefore the number of independent parameters in $\rho$) grows as $d^2-1$.  In order to keep the RMS error ($\Delta_2 = \sqrt{\Tr\left[(\rhotomo-\rho)^2\right]}$) fixed, $N$ must grow proportional to $d$.  Second, $\rhotomo$ has more eigenvalues, so the probability of at least one negative eigenvalue grows with $d$ (for fixed $\Delta_2$).  Together, these scalings ensure that tomographic estimates of large systems are rarely non-negative.

The problems with tomography are well known -- negative eigenvalues were precisely the embarrassing feature that motivated MLE.  As we shall see, however, MLE's implausible zero eigenvalues are closely related to tomography's negative ones.

\subsubsection{How MLE works} \label{secMLEWorkingsMLE}

MLE, though sometimes complex in implementation, is very simple in theory.  Given a measurement record $\M = \left\{M_1,M_2,M_3\ldots M_N\right\}$ (where $M_i$ is a positive operator representing the $i$th observation), the estimator seeks the maximum of the likelihood function,
\begin{equation}
\Lik(\rhohat) = p(\M | \rhohat) = \prod_i{\left(\Tr[M_i\rhohat]\right)}.
\end{equation}

$\M$ can be compactly represented as a list of frequencies.  Define a set $\P = \{E_1\ldots E_m\}$ containing all possible outcomes, and let $n_i$ be the number of times that $E_i$ appears in $\M$.  Then $\M \sim \{n_1\ldots n_m\}$.  As $N$ increases, the frequency representation of $\M$ remains short.

Finding $\rhoMLE$ is feasible because $\Lik(\rhohat)$ has two convenient properties.  First, it is non-negative, so we can maximize $\log(\Lik(\rhohat))$.  Second, $\log(\Lik(\rhohat))$ is convex.  The proof is quite simple: we observe that $\log(\Lik(\rhohat)) = \sum_i{\log\Tr[M_i\rhohat]}$; that $\Tr[M_i\rhohat]$ is a non-negative, linear function of $\rhohat$; that the logarithm of a linear function is convex; and that the sum of convex functions is convex.  Among other things, this means that $\Lik(\rhohat)$ has a unique local maximum.

\subsubsection{The relationship between tomography and MLE} \label{secMLEWorkingsConnection}

The likelihood function has another elegant property:  \textbf{If there is a state $\rhotomo$, such that the probability predicted for every outcome is equal to its observed frequency, then $\rhotomo$ is the maximum of $\Lik(\rhohat)$}.  To prove this, let us write $\log\Lik(\rhohat)$ in terms of (a) the \emph{observed} frequencies ($f_j = \frac{n_j}{N}$), and (b) the \emph{predicted} probabilities ($\phat_j = \Tr[E_j\rhohat]$) for all the $E_j$:
\begin{eqnarray}
\Lik(\rhohat) &=& \prod_i{\left(\Tr[M_i\rhohat]\right)} 
  = \prod_j{\Tr[E_j\rhohat]^{n_j}} \\
\log(\Lik(\rhohat)) &=& \sum_j{n_j\log\left(\Tr[E_j\rhohat]\right)} \\
&=& N\sum_j{f_j\log \phat_j} \\
&=& N\sum_j{\left[f_j\log f_j - (f_j\log f_j-f_j\log \phat_j)\right]} \nonumber \\
&=& -N\left[ H(f) + D(f||\phat) \right].
\end{eqnarray}
The last line invokes two information-theoretic quantities, \emph{entropy} $H(\cdot)$ and \emph{relative entropy} $D(\cdot||\cdot)$.  $H(f)$ doesn't depend on $\phat$, so it is irrelevant for maximization.  The relevant quantity is $D(f||\phat)$, which is always non-negative, and uniquely zero when $\phat=f$.  Thus, $\log(\Lik(\rho))$ is uniquely maximized when $\phat=f$. \qed

So, if $\rhotomo$ is a valid state, then $\rhoMLE=\rhotomo$.  What if $\rhotomo$ exists, but is \emph{not} a valid state?  It must still be Hermitian and have unit trace.  Furthermore, it predicts non-negative probability for each $M_i$ observed, so $\Tr[E_i\rhotomo]\geq0$ for all $i$.  The hyperplanes $\Tr[E_i\rhohat]=0$ define a polytope in Hilbert-Schmidt space -- a simple example is the ``Bloch cube'' referred to previously -- which contains $\rhotomo$.

If we extend the domain of $\Lik(\rhohat)$ to this polytope and its interior, then its maximum must coincide with $\rhotomo$, since $\rhotomo$ predicts the correct frequencies.  Tomography, in other words, is essentially unconstrained MLE.

\begin{figure}[tb]
\includegraphics[width=\FCW]{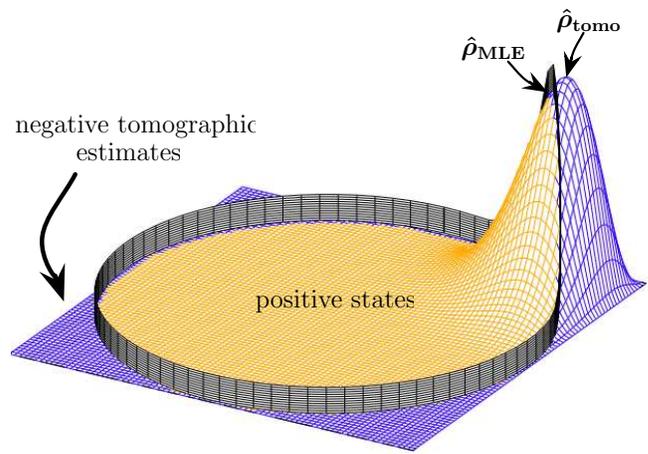}
\caption{An example of a likelihood function (for a single qubit) whose \emph{unconstrained} maximum lies outside the state-set, and whose \emph{constrained} maximum therefore lies on its boundary.  The domain shown here is a cross section of the Bloch sphere, with $\expect{\sigma_y}=0$.  This particular likelihood function is obtained from 16 measurements each of $\sigma_x$ and $\sigma_z$, comprising 14 $\ket{1}$ and $\ket{+}$ results, and 
2 $\ket{0}$ and $\ket{-}$ results.  The unconstrained maximum of $\Lik(\rho)$ is at $\rhotomo = \frac12\left(\Id + \frac34\sigma_x+\frac34\sigma_z\right)$, which has a negative eigenvalue.  The constrained maximum is at $\rhoMLE = \proj{\psi}$, where $\ket{\psi} = \frac{1}{\sqrt{2+\sqrt{2}}}\left( \ket{1}+\ket{+}\right)$.}
\label{figBadLikelihood}
\end{figure}

Because $\Lik(\rhohat)$ has a unique local maximum at $\rhotomo$, its maximum over a closed region which does not contain $\rhotomo$ must lie on the boundary of that region (see Fig. \ref{figBadLikelihood}).  The set of non-negative density matrices is precisely such a closed region, so whenever $\rhotomo$ is not a valid state, $\rhoMLE$ must lie on the boundary of the state-set. That is, it will be rank-deficient.

MLE and tomography are thus variants of the same procedure, distinguished only by the positivity constraint \footnote{In fact, tomography does have a de facto positivity constraint; all the probabilities for observed events must be non-negative.  Quantum mechanics, on the other hand, demands that all the probabilities for \emph{any event that could ever be observed} must be non-negative.  These distinct constraints lead, respectively, to the ``Bloch polytope'' and to the Bloch sphere, as the set of valid states.  This distinction between observ\emph{ed} and observ\emph{able} events is what undermines frequentism in quantum estimation.}.
MLE is a sort of minimal fix for tomography, returning the non-negative state that is in some sense ``closest'' to $\rhotomo$.  Actually computing the \emph{number} of zero eigenvalues in $\rhoMLE$ seems difficult, but numerical exploration for 1, 2, 3, and 4 qubit problems \cite{MalhotraInPrep} suggests that $\rhoMLE$ usually has at least as many zero eigenvalues as $\rhotomo$ has negative ones.  In conjunction with the observation that large-system tomography tends to yield many negative eigenvalues, this explains the many zero eigenvalues in experimental applications of MLE.

\subsection{What is the underlying flaw?} \label{secMLEFlaw}

Tomography and MLE maximize $\Lik(\rhohat)$, over different domains.  They display the same pathology, implying unjustifiable (zero or negative) probabilities.  The underlying problem is simple: \textbf{maximum likelihood methods are \emph{frequentistic}; they interpret observed frequencies as probabilities.}  By maximizing $\Lik(\rhohat)$, they seek to fit the observed frequencies as precisely as possible.  If there exists a $\rhohat$ that fits the data exactly, then that is always the best estimate.

The point of \emph{state} estimation, however, is not solely to explain the data.  Rather, it is to find a state that will predict the future.  It should concisely describe what the estimator knows about the system being estimated.  Mindless data fitting accomplishes only \emph{retro}diction, of the past.  The best description of the past (i.e., data) probably does not describe the estimator's knowledge, especially her uncertainty.

For example, consider estimating the bias of a coin after flipping it just once.  The best fit to the data is to assume that the coin \emph{always} comes up the same way.  This clearly does not describe the estimator's knowledge -- an honest description would perhaps include the word ``scant''.  Ironically, it is the \emph{high} entropy of the estimator's knowledge that causes a spuriously low-entropy estimate.

The problem with MLE is that it matches probabilities to observed frequencies, consistent with \emph{frequentist} statistics.  This is actually unfair to frequentism, which begins by defining probability as the infinite-sample-size limit of frequency.  A true frequentist avoids making statements about probabilities in the absence of an infinite ensemble, so applying a frequentist method to relatively small amounts of data is inherently disaster-prone.  Nonetheless, this is precisely what is happening in MLE.  For further discussion, see Section \ref{secBMEFoundation}.

\section{Bayesian Mean Estimation} \label{secBME}

Bayesian methods provide a different perspective on statistics.  The procedure presented here, Bayesian mean estimation (BME), avoids the pitfalls of MLE. Here are three basic tenets, each of which independently motivates BME:
\begin{enumerate}
\item ``Consider all the possibilities.''  MLE identifies the \emph{best} fit to observed data, but many nearby states are almost equally likely.  An honest estimate should incorporate these alternatives.
\item ``Demand error bars.''  The estimate should be compatible with error bars, e.g. $\rhohat\pm\Delta\rho$.  This implies a ball containing most of the plausible states, of size $\Delta\rho$, with $\rhohat$ somewhere around the center.  If $\rhohat$ is rank-deficient, no such region exists.  Thus, $\rhohat$ should lie far enough from the state-set's boundary to be compatible with well-motivated error bars.
\item ``Optimize accuracy.'' Obviously, the estimate $\rhohat$ should be close to the true $\rho$.  How do we evaluate this?  Quantum strictly proper scoring rules \cite{RBK06} yield a class of metrics designed to measure this closeness, called \emph{operational divergences}.  BME uniquely minimizes the expected value of every operational divergence.
\end{enumerate}
Each of these motivations illustrates one of BME's major advantages.  The estimate predicts reliable probabilities for all measurement outcomes, it comes with a free set of error bars, and it is (on average) the most accurate estimate that can be made from the data.

Bayesian approaches have been previously discussed in various contexts.  Helstrom \cite{HelstromBook76} applied Bayesian methodology extensively to estimation.  He considered a variety of utility functions, especially the rather pathological $\delta$-function utility that motivates MLE, without paying particular attention to the posterior mean.  Jones \cite{JonesAOP91} applied Bayesian inference with Haar measure, focusing on information-theoretic bounds.  Derka et al \cite{DerkaJFMO96} examined Bayesian estimation in some detail, primarily in its connections to tomography and maximum-entropy.  Schack et al \cite{SchackPRA01} formalized a deep connection to exchangeable (deFinetti) states.  More recently, Neri \cite{Neri05} considered Bayesian estimation of phase difference in coherent light.  Tanaka and Komaki \cite{TanakaPRA05} proved the optimality of Bayesian estimation with respect to relative entropy.

My goal in this section is to propose BME as a \emph{practical} procedure for state estimation, and to describe its operational advantages.  I begin by concisely presenting the BME algorithm, then discuss in Section \ref{secBMEMetropolis} how it can be implemented.  Section \ref{secBMEProperties} analyzes the properties of $\rhoBME$, focusing on the three advantages asserted above.  Finally, Section \ref{secBMEFoundation} contrasts the Bayesian and frequentist approaches.

\subsection{The BME Algorithm} \label{secBMEAlgorithm}

Bayesian mean estimation is conceptually simple.
\begin{enumerate}
\item Use the data to generate a likelihood function, $\Lik(\rhohat) = p(\M|\rhohat)$.  $\Lik$ is not a probability distribution; it quantifies the \emph{relative} plausibility of different state assignments.
\item Choose a \emph{prior distribution} over states, $\pi_0(\rhohat)\diff\rhohat$.  It represents the estimator's ignorance, and should generally be chosen to be as ``uniform'', or uninformative, as possible.
\item Multiply the prior by the likelihood, and normalize to obtain a posterior distribution
\begin{equation}
\pi_f(\rhohat)\diff\rhohat \propto \Lik(\rhohat)\pi_0(\rhohat)\diff\rhohat,
\end{equation}
which represents the estimator's knowledge. The proportionality constant is set by normalization.
\item Report the mean of this posterior,
\begin{equation}
\rhoBME = \int{\rhohat\pi_f(\rho)\diff\rhohat}.
\end{equation}
This is the best concise description of the estimator's knowledge.
\end{enumerate}

\subsection{Implementation} \label{secBMEMetropolis}

In practice, BME comes down to computing an integral.  The \emph{best} way of doing this remains uncertain, as does the existence of an exact solution.  The numerical algorithm presented below has been demonstrated to work well in a small variety of cases.  However, it could be improved in many ways, and has some glaring deficiences.  This algorithm should thus be taken as a proof of principle (i.e., it's \emph{possible} to do Bayesian estimation) rather than an optimal approach.

An important observation for \emph{any} integration procedure is that the likelihood is easy to compute.  $\Lik(\rho)$ is the probability of observing a sequence of outcomes $\M = \{M_1\ldots M_N\}$, given $\rho$.  This is the product of the probabilities for the individual $M_i$, each of which is given by Born's rule:
\begin{equation}
\Lik(\rho) = \Tr(M_1\rho)\Tr(M_2\rho)\Tr(M_3\rho)\ldots\Tr(M_N\rho)
\label{eqLikelihoodProdForm}
\end{equation}
When $\M$ is represented using frequencies ($E_i$ was observed $n_i$ times, for $i\in[1\ldots m]$), this can be evaluated in $O(m)$ time:
\begin{equation}
\Lik(\rho) = \Tr(E_1\rho)^n_1\Tr(E_2\rho)^n_2\ldots\Tr(E_m\rho)^n_m.
\end{equation}

\subsubsection{The Metropolis-Hastings algorithm} \label{secBMEMetropolisGeneral}

In the absence of an analytic solution to the integral, we fall back to numerical Monte Carlo methods. Because $\Lik(\rho)$ is usually a sharply peaked function over a high-dimensional space, brute-force random sampling will converge extremely slowly.  Metropolis algorithms \cite{MetropolisJCP53} were conceived for precisely such situations.  A variant known as \emph{Metropolis-Hastings} \cite{HastingsBiometrika70,ChibTAS95} is commonly used for Bayesian estimation, and can be adapted straightforwardly to quantum states.

The Metropolis-Hastings algorithm computes the average value of a function (in this case, $\rho$) over an integration measure (in this case, $\Lik(\rho)\pi_0(\rho)\diff\rho$).  It leverages the fact that $\Lik(\rho)\pi_0(\rho)\diff\rho$ is typically dominated by a small region of high likelihood.  Whereas basic Monte Carlo methods jump randomly around the integration measure, Metropolis-Hastings makes local, biased jumps.  This samples the most relevant parts of the sample space preferentially.  After each jump, the current value of $\rho$ is added to a running tally.  This tally, divided by the total number of jumps, becomes the final average.

To implement Metropolis-Hastings, we begin with a rule $J$ for jumping from any $\rho$ to a nearby $\rho' = J(\rho)$.  The precise form of the rule is unimportant; it is usually stochastic, although a deterministic rule (traversing a quasi-random set) is conceivable.  What is important is that $J$ should generate the underlying measure $\diff\rho$: for any $\rho_0$, the set $\{J^n(\rho_0): n\in[0\ldots N]\}$ should sample uniformly from $\diff\rho$ as $N\rightarrow\infty$.  For example, the following rule implements Lebesgue measure over the interval $[0\ldots1]$: $J(x) = (x + y)\ \mathrm{mod}\ 1$, where $y$ is selected from a Gaussian distribution with zero mean and fixed variance.

Such a rule, unmodified, would compute $\int_\rho{f(\rho)\diff\rho}$.  To average instead over $\Lik(\rho)\pi_0(\rho)\diff\rho$, we modify the rule as follows.  After choosing $\rho'$, but before jumping to it, we compute the likelihood ratio
\begin{equation}
r = \frac{\Lik(\rho')\pi_0(\rho')}{\Lik(\rho)\pi_0(\rho)}.
\end{equation}
If $r>1$ ($\rho'$ is more likely than $\rho$) then we jump as before.  If not, then we jump to $\rho'$ with probability $r$, and stay at $\rho$ (adding it, once again, to the running total) with probability $1-r$.

This biasing ensures that the algorithm spends more time at more likely spots, and tends to lurch uphill into regions of high probability.  Unlike a gradient algorithm (as might be used for MLE), it does not actively seek the point of highest probability; jumping to a region of lower probability is both possible and necessary.  Detailed discussion and explanation of why this works can be found in \cite{ChibTAS95}.

\subsubsection{Applying Metropolis-Hastings to quantum states} \label{secBMEMetropolisDetails}

The heart of the algorithm is the rule $J$.  It determines $\diff\rho$, and its form is critical to the algorithm's performance.  Different underlying measures will require different rules.  Measures with some claim to ``uniformity'' are usually invariant under a symmmetry group.  The natural group for quantum states on a $d$-dimensional Hilbert space is $SU(d)$, and the measure that this group induces over pure states is called \emph{Haar measure}.

A sensible prior should extend over \emph{all} possible states, so we need measures extending to mixed states.  However, there is no uniquely suitable measure over mixed states, because their spectral degrees of freedom (eigenvalues) have no obvious symmetry.  One appealing class of measures, proposed by Zyzkowski and Sommers \cite{ZyczkowskiJPA01}, is the set of \emph{induced measures}, denoted here by $\diffk{k}\rho$.  They are obtained by beginning with Haar measure on a $d\times k$ dimensional system, then tracing out the ancillary factor.  Thus, $\diffk{1}\rho$ is simply the Haar measure on pure states; while $\diffk{d}\rho$ is the Hilbert-Schmidt measure (Lebesgue measure on the vector space of Hermitian $d\times d$ matrices).

These induced measures are easy to implement.  Instead of keeping track of $\rho$ itself, we generate and track a pure state $\ket{\psi_{d\times k}}$ in $d\times k$ dimensions.  At each step, $\rho$ is obtained by tracing out part of $\proj{\psi_{d\times k}}$.  The ancillary degree of freedom acts as a sort of hidden variable, internal to the algorithm.  We need only a rule $J$ to implement Haar measure over the larger Hilbert space.

This could be done in many ways -- for instance, at each step, we could generate a random unitary from Haar measure.  This has two huge drawbacks.  First, the jumps are nonlocal, which negates the key advantage of the Metropolis-Hastings algorithm.  Second, generating and applying a random unitary is computationally expensive.  Instead, we need a relatively small set of efficiently constructable unitaries that generate the entire group.

Here is an efficient local random walk rule that generates Haar measure on a $d$-dimensional Hilbert space:
\begin{enumerate}
\item Choose a \emph{direction}, by generating two random integers $i,j\in[0\ldots d-1]$.  Select a Hermitian operator $H_{ij}$ that acts only on the $\{\ket{i},\ket{j}\}$ subspace.  Define $H_{ij}=\sigma_z$ if $i=j$, $H=\sigma_x$ if $i<j$, and $H=\sigma_y$ if $i>j$.
\item Choose a \emph{distance}, $\delta$, from a distribution (e.g., Gaussian) with $\expect{\delta}=0$ and $\expect{\delta^2}=\Delta^2$.  We will discuss the choice of $\Delta$ below.
\item Let $J(\ket{\psi}) = e^{iH_{ij}}\delta\ket{\psi}$.  Since $U$ acts nontrivially only on the $\ket{i},\ket{j}$ subspace, this can be done very easily and quickly.
\end{enumerate}
Each step's distance is chosen randomly to ensure uniform sampling -- with a fixed stepsize, it is just barely conceivable that this algorithm might trace out a discrete lattice of states.  The \emph{average} step size $\Delta$ is important:  if $\Delta$ is too large, the algorithm will not find small regions of high probability efficiently; if $\Delta$ is too small, it will explore the space very slowly.  The optimal $\Delta$ will depend on $\Lik(\rho)$, and there is no way to identify it a priori.

The algorithm must therefore vary $\Delta$ dynamically, with feedback.  If $\Delta$ is very, very small, then almost every jump will be accepted, whereas if $\Delta$ is large, very few will be.  A good heuristic is that the acceptance ratio should be around $60\%$ \cite{WikiMetropolisHastings} (other values are also suggested \cite{ChibTAS95}).  The algorithm should track the acceptance ratio over the last $\sim\!1000$ jumps, gradually changing $\Delta$ as appropriate to maintain it around $60\%$.  

Dynamically adjusting the step size like this can, in theory, break the convergence properties of the algorithm.  This occurs if the distribution over states is multimodal; the step size is reduced in order to explore one narrow peak in detail, and a far-off peak becomes inaccessible.  Fortunately, the likelihood function itself is guaranteed to be log-convex, and therefore uni-modal.  For well-behaved priors with convex support (e.g., the Hilbert-Schmidt prior, $\diffk{d}\rho$), this means that $\Lik(\rho)\pi_0(\rho)$ can safely be sampled this way.

Other priors -- in particular, the Haar prior, which is interesting as a limiting case -- do not have convex support.  These priors will yield multimodal posterior distributions.  How to effectively and reliably sample from such distributions is an outstanding problem.  Repeating the sampling many times, with randomly distributed starting points, is not reliable.  It fails badly if two similar peaks in the distribution have unequally sized regions of convergence; the peak with the larger convergence region will be relatively oversampled.

\subsection{[Good] Properties of the BME estimate} \label{secBMEProperties}

Why should an experimentalist use Bayesian mean estimation?  After all, BME (via Monte Carlo) is more computationally intensive than MLE.  The answer, of course, is that BME provides a \emph{better} estimate than MLE.  Specifically: (1) $\rhoBME$'s eigenvalues are never unjustifiably small (or zero); (2) the procedure can easily be made to yield well-motivated error bars that are compatible with $\rhoBME$; (3) BME is, in a particular sense, the most accurate possible estimate -- not just asymptotically, but for finite $N$.

\subsubsection{The estimate is plausible} \label{secBMEPropertiesPlausible}

The first objection to MLE is that $\rhoMLE$ is implausible; it can (and often does) have zero eigenvalues, which imply an unjustified certainty.  Any alternative procedure should yield a strictly positive estimate.  BME yields just such an estimate, subject to a very weak restriction on the prior.

Consider a simple and illustrative example in classical estimation.  We estimate the bias $b$ of a coin, which comes up ``heads'' with probability $b$, and ``tails'' with probability $1-b$.

Flipping the coin $N$ times yields a measurement record consisting of $n$ heads and $N-n$ tails.  The likelihood function is
\begin{equation}
\Lik(b) = b^n(1-b)^{N-n},
\end{equation}
and so the MLE estimate is
\begin{equation}
\hat{b}_{\mathrm{MLE}} = \frac{n}{N}.
\end{equation}
If $n=0$ or $n=N$, then $\hat{b}_{\mathrm{MLE}}$ will assign zero probability to observing either ``heads'' or ``tails'', respectively.

If we adopt a Bayesian approach, then we must choose a prior -- e.g., the uniform prior with respect to Lebesgue measure, $\pi_0(b)\diff b = \diff b$.  The mean of the posterior is an integral of the likelihood, and we get:
\begin{equation}
\hat{b}_{\mathrm{BME}} = \frac{n+1}{N+2}.
\end{equation}
Since $0\leq n \leq N$, the Bayesian estimator never assigns zero probability to anything.  The lowest possible probability assignment for either heads or tails is $p_{\mathrm{min}} = \frac{1}{N+2}$.  With no data at all, the Bayesian assigns $p=\frac12$ to both outcomes; after a single flip, she assigns $p=\frac23$ to the outcome that was observed, and $p=\frac13$ to the other.  

This is the property that we want in a quantum estimation procedure.  The probabilities assigned to unobserved events are not only nonzero, but also sensible -- after $N$ trials, it's reasonable to assume that the probability of an as-yet-unobserved outcome is at most $1/N$, and to assign $p_{\mathrm{min}} \approx \frac1N$.

However, this property depends on the prior.  Consider the prior $\pi_0(b) = \frac12\left(\delta(b)+\delta(1-b)\right)$.  After one observation of ``tails,'' our Bayesian estimate would be $\hat{b}_{\mathrm{BME}} = 0$, which is implausible.  The problem is that a finite number of observations (one) ruled out every $b$ in the support of $\pi_0$ that ascribed nonzero probability to ``heads.''

The situation gets even worse if the \emph{next} observation is ``heads.''  The data now rule out every hypothesis, the posterior $\pi_f(\rho)\diff\rho$ vanishes entirely, and the Bayesian procedure simply fails.  This stems from a contradiction.  A prior over \emph{states} implies a probability distribution over \emph{observations} as well.  $\pi_0$ assigned exactly zero probability to $\M=\{\mathrm{``heads''},\mathrm{``tails''}\}$ -- which was then observed, causing a contradiction.

The following statements about a prior $\pi_0$ are logically equivalent:
\begin{enumerate}[(a)]
\item $\pi_0$ assigns zero probability to some [finite-length] measurement record.
\item Bayesian estimation using $\pi_0$ will, for some measurement record, yield an estimate with a zero probability.
\item There exists a measurement record that will annihilate $\pi_0$, so that Bayesian estimation fails completely.
\end{enumerate}
Let us define a \textbf{fragile} prior as one for which these statements hold (and which can therefore yield a rank-deficient estimate).  An estimator should choose a \textbf{robust} (i.e., not fragile) prior, which in turn guarantees a full-rank estimate.

In classical probability estimation, avoiding fragility is simple: \emph{A prior is robust if and only if it has support on the interior of the probability simplex.}  States in the interior do not predict zero probability for any observation.  They can never be ruled out, so a prior supported on one can never be annihilated by the data.  Conversely, every prior supported only on the boundary will be annihilated by a measurement record that includes every possible outcome.

Intriguingly, this condition does not extend to the quantum problem.  Support on the interior (i.e., on the full-rank states) is sufficient, but not necessary, for robustness.  Consider estimation of a single qubit using the Haar prior, which is restricted to (and uniform over) the pure states.  Each observation rules out, at most, a single pure state -- if $\proj{0}$ is observed, then the true state cannot be $\proj{1}$.  There are uncountably many distinct candidate pure states, which means that no [finite-length] measurement record can annihilate the prior.  The Haar prior is robust.

As a general rule, just about every prior that a halfway-sane estimator would pick is, in fact, robust.  Not only the Haar prior (which implies absolute certainty that $\rho$ is pure), but much more extreme priors, such as an equatorial distribution on the Bloch sphere -- or, for that matter, any continuous curve on the Bloch sphere's surface -- are robust.  Appendix \ref{appPlausibilityTheorem} demonstrates a necessary and sufficient condition.  

\subsubsection{The estimate comes with natural error bars} \label{secBMEPropertiesErrorbars}

Another objection to the MLE procedure is that $\rhoMLE$ is not, in general, compatible with any error bars.  This is an obvious consequence of zero eigenvalues; error bars imply a region of plausibility \emph{surrounding} the point estimate.  When the estimate lies on the state-set's boundary, no such region can exist -- in order that $\rhoMLE$ be in its interior, the region would have to contain negative matrices.

The BME estimate is always full-rank, which is encouraging.  This in itself does not guarantee compatibility with sensible error bars.  The estimate $\hat{\rho} = \frac{99}{100}\rhoMLE + \frac{1}{100d}\Id$ is full-rank, but the estimator's honest uncertainty about $\hat{\rho}$ might well be greater than $\pm1\%$.  Happily, the BME estimation procedure can easily be adapted to yield natural error bars, which are compatible with the point estimate.

First, let us consider what form these error bars should take.  Intuitively, the qualified estimate should look like
\begin{equation}
\rho = \hat{\rho} \pm \Delta\rho,
\end{equation}
but what, precisely, is ``$\Delta\rho$''?  As $\hat{\rho}$ is a $d\times d$ matrix, we might suppose that $\Delta\rho$ is also a $d\times d$ matrix, so $\rho_{ij} = \hat{\rho}_{ij}\pm \Delta\rho_{ij}$.  This fails to account for \emph{covariance} between distinct elements of $\rho$.  For example, the diagonal elements of $\rhohat$ must vary together to maintain $\Tr(\rho) = 1$.

The correct way to think about the estimator's uncertainty begins by representing the estimate, $\rhoBME$, as a $d^2-1$ dimensional vector in Hilbert-Schmidt space.  For a single qubit:
\begin{equation}
\rho \sim \left(\begin{array}{c}x\\ y \\ z\end{array}\right)
\equiv \left(\begin{array}{c}\Tr(\sigma_x\rho)\\ \Tr(\sigma_y\rho) \\ \Tr(\sigma_z\rho)\end{array}\right).
\end{equation}
The estimator's uncertainty is represented as a symmetric covariance matrix on the same space:
\begin{equation}
\Delta\rho \sim \left(\begin{array}{ccc}
\Delta x^2\ &\ \Delta xy\ &\ \Delta xz \\
\Delta xy\ &\ \Delta y^2\ &\ \Delta yz \\
\Delta xz\ &\ \Delta yz\ &\ \Delta z^2
\end{array}\right)
\end{equation}
The elements of $\Delta\rho$ involve two different expectation values: one with respect to the state, denoted $\expect{X}_\rho \equiv \Tr(X\rho)$; and one with respect to the posterior probability, denoted $\overline{f} \equiv \int{f(\rho)\pi(\rho)\diff\rho}$.  Using this notation,
\begin{eqnarray}
\Delta x^2 = \overline{\expect{x}^2}-\left(\overline{\expect{x}}\right)^2,
\end{eqnarray}
with the other elements given by the obvious generalization.

Represented as a covariance matrix, $\Delta\rho$ quantifies the second cumulants of the estimator's probability distribution $\pi_f(\rho)\diff\rho$.  It defines an ellipsoid in Hilbert-Schmidt space, which is a \emph{credible interval} (the Bayesian version of a confidence interval).  The eigenvectors of $\Delta\rho$ are operators that define the principal axes of this ellipsoid, and the corresponding eigenvalues are their lengths.

As a matrix that acts on density matrices, $\Delta\rho$ is a superoperator.  It is symmetric and non-negative, but not completely positive or trace-preserving, so it cannot be interpreted as a quantum process.  However, the superoperator interpretation gives a formula for the estimator's uncertainty about the expectation value of a particular operator $X$.  Defining $\Delta\rho[X]$ to be the superoperator's action on $X$,
\begin{equation}
\Delta\expect{X}^2 = \Tr\left( X^\dagger\ \Delta\rho[X] \right)
\end{equation}
quantifies the estimator's expected error in $\expect{X}$.

Alternatively, $\Delta\rho$ can be represented as an unnormalized symmetric bipartite state,
\begin{eqnarray}
\Delta\rho &=& \overline{\rho\otimes\rho} - \overline{\rho}\otimes\overline{\rho} \\
&=& \int{\rho\otimes\rho\ \pi_f(\rho)\diff\rho} - \rhoBME\otimes\rhoBME,
\end{eqnarray}
and in this representation, the estimator's expected error in $\expect{X}$ is
\begin{equation}
\Delta\expect{X}^2 = \Tr\left(X\otimes X \Delta\rho\right).
\end{equation}

This $\Delta\rho$ is a consistent representation of the estimator's uncertainty; for any $X$, it yields the same $\Delta\expect{X}^2$ that an independent estimate of $\expect{X}$ would.  To see this, let $X$ be an arbitrary observable with eigenvalues between $x_\mathrm{min}$ and $x_\mathrm{max}$.  The variance computed via BME is:
\begin{eqnarray}
\Delta\expect{X}^2 &=& \Tr(X\otimes X\Delta\rho) \nonumber \\
&=& \Tr\left[ X\otimes X \int{\!\rho\otimes\rho\ \pi_f(\rho)\diff\rho}\right] \nonumber \\
&& - \Tr\left[X\otimes X \rhoBME\otimes\rhoBME\right] \nonumber \\
&=& \int{\!\Tr[X\rho]\cdot\Tr[X\rho]\pi_f(\rho)\diff\rho} - \Tr[X\rhoBME]^2 \nonumber \\
&=& \int{\!\expect{X}_\rho^2\pi_f(\rho)\diff\rho} - \left[\int{\!\expect{X}_\rho\pi_f(\rho)\diff\rho}\right]^2
\end{eqnarray}
Because $\expect{X}$ parameterizes exactly one of the dimensions of Hilbert-Schmidt space, we can compute a marginal distribution over $\expect{X}$ by integrating $\pi_f(\rho)\diff\rho$ over its other $d^2-2$ dimensions, which we denote by $\sigma$.  Then $\diff\rho = \diff{\expect{X}}\diff\sigma$, and
\begin{equation}
\pi_f(\expect{X})\diff\expect{X} \equiv \int_\sigma{\pi_f(\rho)\diff\rho},
\end{equation}
in terms of which,
\begin{equation}
\Delta\expect{X}^2 = \int{ \expect{X}^2\pi_f(\expect{X})\diff\expect{X}} - \left[\int{\expect{X}\pi_f(\expect{X})\diff\expect{X}}\right]^2,
\end{equation}
which is the familiar formula for the variance of the univariate distribution $\pi_f(\expect{X})$.

In particular, if $\ket{\psi}$ is an eigenvector of $\rhoBME$, let $X = \proj{\psi}$.  Then $\lambda = \expect{X}$ is the corresponding eigenvalue, and $\Delta\lambda^2 = \Delta\expect{X}^2$ is the reported uncertainty about it.  Since $\expect{X}_\rho$ is between 0 and 1 for all $\rho$, $\pi_f(\lambda)\diff\lambda$ is a distribution over the interval $[0\ldots1]$.  For any such distribution, $\Delta\lambda^2\leq\lambda(1-\lambda)$, so every eigenvalue yields an upper bound for its own uncertainty.  Note, too, that this bound is uniquely saturated by $\pi(\lambda) = (1-p)\delta(\lambda) + p\delta(1-\lambda)$, which is maximally bimodal.  In practice, well-behaved priors will produce convex posteriors, for which $\Delta\lambda^2\lesssim\lambda^2$ (i.e., $\Delta\lambda$ is no greater than $\lambda$ itself) can reasonably be expected.

\subsubsection{BME optimizes accuracy} \label{secBMEPropertiesAccuracy}

Above all else, an estimation procedure should yield an \emph{accurate} estimate -- one as close to the ``true'' state as possible.  While the concept of a ``true'' state is problematic in actual experiments, it makes perfect sense in the context of a simple game.  An impartial judge selects a state $\rho$, and provides $N$ copies of it to the estimator, who measures them and reports an estimate $\rhohat$.  The best procedure is the one that consistently makes $\rhohat$ as close as possible to the unknown [to the estimator] $\rho$.

The point of this section is to show that BME is the \emph{most} accurate scheme possible, in the sense that the \emph{expected} error between $\rhohat$ and $\rho$ is minimal.  The argument presented here is brief; for more detail see \cite{RBK06}.  This optimality holds for every finite $N$, not just asymptotically.  It depends, of course, on the measure of ``error'' between $\rho$ and $\rhohat$ adopted.  The error measures optimized by BME, \emph{operational divergences}, are arguably the best-motivated such measures.

Operational divergences, denoted $\Delta(\rho:\rhohat)$, measure how well the density matrix $\rhohat$ describes (or estimates) the quantum state $\rho$.  A certain subtlety should be noted here:  whereas $\rho$ represents the state of a quantum system, $\rhohat$ is a classical description of a state -- e.g., a density matrix written down on paper.  Two natural requirements constrain operational divergences.  First, $\Delta$ must represent the outcome of some physically implementable process.  Second, the best description of $\rho$ had better be $\rho$ itself.

To satisfy operationality, we imagine trying to motivate the estimator to do a good job.  A third-party verifier, equipped with the estimate $\rhohat$, will perform a measurement on $\rho$.  This measurement, $\P(\rhohat) = \{E_1\ldots E_m\}$, is an arbitrary POVM that may depend on $\rhohat$.  Depending on the outcome ($i$), the verifier pays the estimator an amount $r_i(\rhohat)$.  

The estimator's reward is represented by an operator
\begin{equation}
\R(\rhohat) = \sum_i{r_i(\rhohat)E_i(\rhohat)},
\end{equation}
and her expected reward (which she hopes to maximize) is
\begin{equation}
r(\rho:\rhohat) = \Tr(\rho\R(\rhohat)), \label{eqRLinearity}
\end{equation}
The amount that she loses by inaccurately describing the state,
\begin{equation}
\Delta(\rho:\rhohat) \equiv r(\rho:\rhohat)-r(\rho:\rho) = \Tr\left[\rho\left(\R(\rhohat)-\R(\rho)\right)\right].
\end{equation}
is an operational divergence.  Note that: (1) it is operationally significant; and (2) the best description of $\rho$ is $\rho$ itself.

Of course, not every reward scheme is \emph{strictly proper}, satisfying the condition that $\rho$ be its own best estimate,
\begin{equation}
r(\rho:\rho) > r(\rho:\rhohat)\ \forall\ \rhohat\neq\rho.
\label{eqStrictPropriety}
\end{equation}
Equation \ref{eqStrictPropriety} is a constraint on $\R(\rhohat)$.  If we define $G(\rho) \equiv r(\rho:\rho)$ as the expected reward for a perfect estimate, then a bit of algebra yields
\begin{equation}
G(\rho) > G(\rhohat) + \Tr\left[\left(\rho-\rhohat\right)\R(\rhohat)\right].
\label{eqRewardConstraint}
\end{equation}
Eq. \ref{eqRewardConstraint} holds if and only if: (1) $r(\rho:\rhohat)$ (as a function of $\rho$) is tangent to $G(\rho)$; and (2) $G(\cdot)$ is strictly concave.  Thus, for for every strictly concave function $G(\cdot)$ on density operators, there is a unique operational divergence \footnote{Actually, if $G(\cdot)$ is not differentiable at a point (i.e., it has a cusp), then a family of operational divergences exist, indexed by the possible subgradients $\nabla G(\cdot)$.  This seems to be a purely technical point, with no real significance in practice.}:
\begin{equation}
\Delta(\rho:\rhohat) = G(\rho)-G(\rhohat) - \Tr\left[(\rho-\rhohat)\nabla G(\rhohat)\right],
\label{eqOpDiv}
\end{equation}
where $\nabla G(\cdot)$ is the gradient of $G(\cdot)$.

Operational divergences include widely used measures such as the squared Hilbert-Schmidt or $L_2$ distance,
\begin{equation}
\Delta_2(\rho:\rhohat) = \Tr\left[(\rho-\rhohat)^2\right],
\end{equation}
associated with $G(\rho)=\Tr(\rho^2)$; and the relative entropy or Kullback-Leibler divergence,
\begin{equation}
\Delta_{KL}(\rho:\rhohat) = \Tr\left[\rho\left(\log\rho-\log\rhohat\right)\right],
\end{equation}
associated with $G(\rho) = -H(\rho) = \Tr(\rho\log\rho)$.

Now that we have determined how to measure accuracy, let's try to optimize it.  This is an easy task for an omniscient estimator, because the best estimate of $\rho$ is $\rho$ itself.  If the estimator actually knows $\rho$, then her best plan is to report $\rhohat=\rho$.  The interesting case is an \emph{uncertain} estimator.  She must consider all the \emph{possible} $\rho$, in order to choose the best $\rhohat$.  A risk-neutral estimator seeks to maximize her expected reward, averaged over all possible $\rho$.

Consider \emph{any} estimation procedure, as a map from measurement records $\M$ to estimates $\rhohat(\M)$.  Which procedure should the estimator choose?  Suppose that the unknown state $\rho$ to be estimated will be drawn from an ensemble described by $\pi_0(\rho)\diff\rho$.  The \emph{expected} reward yielded by the procedure $\rhohat(\M)$ is an average over (a) possible $\rho$, and (b) the ensuing $\M$.
\begin{equation}
\overline{r} = \int_\rho{\pi_0(\rho)\diff\rho \sum_\M{p(\M|\rho)r\left(\rho:\rhohat(\M)\right)}}
\end{equation}
Inserting $r(\rho:\rhohat) = \Tr\left[\rho\R(\rhohat)\right]$ (Eq. \ref{eqRLinearity}),
\begin{equation}
\overline{r} = \int_\rho{\pi_0(\rho)\diff\rho \sum_\M{p(\M|\rho)\Tr\left[\rho\R\left(\rhohat(\M)\right)\right]}}.
\end{equation}
The trace, sum, and integral are all linear, so we can rearrange them as
\begin{equation}
\overline{r} = \sum_\M{
  \Tr\left[
    \left(
      \int_\rho{ \rho p(\M|\rho)\pi_0(\rho)\diff\rho }
    \right)\R\left(\rhohat(\M)\right)
  \right] }.
\end{equation}
We now observe that $\int{p(\M|\rho)\pi_0(\rho)\diff\rho}=p(\M)$, the unconditional probability of observing $\M$.  Furthermore, $\int{\rho p(\M|\rho)\pi_0(\rho)\diff\rho}=\rhoBME(\M)$, the BME estimate given $\pi_0$.  Using these identities, the estimator's expected reward is
\begin{eqnarray}
\overline{r} &=& \sum_\M{p(\M)\Tr\left[\rhoBME(\M)\R\left(\rhohat(\M)\right)\right]} \\
&=& \sum_\M{p(\M)r\left(\rhoBME(\M):\rhohat(\M)\right)},
\end{eqnarray}
and each term in the sum can be independently maximized.  For each $\M$, the optimal $\rhohat(\M)$ is $\rhoBME$ -- which means that BME is unconditionally the optimal estimation procedure.

This result is remarkable because it makes no appeal to asymptotics; the optimality holds for 100, 10, or even just 1 observation.  Of course, when the estimator has insufficient data, the resulting estimate will not be very accurate -- but neither will any other estimate.  Crucially, her uncertainty will be reflected in a highly mixed estimate, with large error bars.  Unlike MLE, BME fails gracefully, making the best use of the available data without over-reaching.

Two points should, however, be kept in mind.  First, BME is not necessarily optimal according to standards that are not operational divergences -- e.g., trace-distance or fidelity.  Measuring the performance of an estimation algorithm by these standards is generally unwise, but they are commonly misused in this way (especially fidelity).  Second, the optimality proof assumes that the estimator's prior coincides with the ensemble from which the unknown states were selected.  A sufficiently wrong prior will lead to horrendous results.  Further research into uninformative priors, and techniques for selecting priors, may alleviate this problem.

\subsection{Bayesian and Frequentist approaches} \label{secBMEFoundation}

Having examined both frequentist and Bayesian approaches, I have focused on the concrete details -- how does MLE fail? why does BME do better? how is BME done? -- because estimation is an operational task.  Certain readers may, however, ask ``What's wrong with frequentism, anyway?''  Others may be wondering what really distinguishes Bayesian and frequentist methods, since $\Lik(\rhohat)$ is crucial to both.  I attempt to address these questions below.

\subsubsection{Why frequentism fails} 

The frequentist approach has dominated statistics for most of the 20th century, so its failure in quantum state estimation requires some explanation.  To see why frequentism fails, we might first ask why it should succeed.

MLE attempts to fit the observed data, and so the MLE estimate is the best ``predictor'' of the past.  Since the goal of a state estimate is to predict the future, frequentist estimation techniques can be justified by the following axiom: \emph{the future will look [statistically] identical to the past}.  If this axiom is true, then $\rhoMLE$ is the best possible estimate.  The Law of Large Numbers implies its validity as $N\rightarrow\infty$, and the Central Limit Theorem quantifies this convergence.

For classical systems, it is always \emph{possible} that the Frequentist Axiom will hold.  If the coin comes up ``heads'' the first time, it's entirely possible that it will always come up heads.  Moreover, the rules are not going to change -- the possible outcomes in the past were ``heads'' and ``tails,'' and they will remain the only possible outcomes in the future.

This doesn't hold for quantum systems.  The past, represented by the estimator's data, comprises a finite set of observations extracted from a finite variety of measurements.  For instance, the estimator might have measured $\sigma_x$, $\sigma_y$, and $\sigma_z$ on a qubit.  Future experimenters, however, might choose to measure \emph{any} observable -- and there are infinitely many.  A quantum state, by definition, predicts the probabilities for every possible measurement.  The frequentist axiom cannot possibly hold; any future observer could violate it at will, simply by making a novel measurement.

Frequentist methods for classical probabilities yield zero probabilities only when
\begin{enumerate}[(a)]
\item event ``$i$'' has never been observed,
\item in every trial, something in the complement of event ``$i$'' \emph{was} observed.
\end{enumerate}
That is, event ``$i$'' \emph{could} have happened, but it didn't.  When MLE is used on quantum systems, the $\proj{\phi_i}$ that ends up getting assigned zero probability is almost never something that could have been observed.  The Achilles' heel of frequentist quantum estimation is that it happily assigns zero probability to events that were never observed \emph{not} to happen.  To avoid this problem, we need a method that does not begin by assuming ``the future will look like the past,'' because for a quantum system, that can't be true.

\subsubsection{How the Bayesian and frequentist approaches differ}

$\Lik(\rhohat)$ is the key ingredient in Bayesian methods, just as in frequentist ones.  It represents everything relevant about the data.  In frequentist methods, $\Lik(\rhohat)$ is the \emph{sole} ingredient, and so the only natural thing to do is to find the $\rhohat$ that maximizes it.  Bayesian methods, in contrast, transform the likelihood into a probability distribution,
\begin{equation}
\Lik(\rhohat) \longrightarrow \pi(\rhohat)\diff\rhohat \propto \Lik(\rhohat)\pi_0(\rhohat)\diff\rhohat,
\end{equation}
by multiplying it by a prior distribution $\pi_0(\rho)\diff\rho$.

A common misconception is that this transformation is trivial when $\pi_0(\rho)\diff\rhohat$ is ``flat'' (e.g., coincides with a Lebesgue or Haar measure).  On the contrary, it transforms a \emph{function} into a \emph{distribution} (or measure), which is an entirely different mathematical object.  Functions, like $\Lik(\rhohat)$, have values.  Distributions have integrals -- they assign values not to points, but to regions.

For example, if $\Lik(x)$ is defined for real-valued $x$, then $\Lik(0)$ and $\Lik(1)$ are well-defined, but $\int_0^1{\Lik(x)}$ is purely meaningless.  To integrate, we must multiply by $\diff x$ (a measure), obtaining a distribution $\Lik(x)\diff x$.  This can be integrated over the interval $[0,1]$ -- but evaluating $\Lik(x)\diff x$ at $x=0$ is ill-defined (and infinitesimal in any case).

This difference between functions and distributions enforces a difference in approach between frequentist and Bayesian methods.  Frequentists, abjuring priors, can only work with the function $\Lik(\rhohat)$.  The corresponding estimate, $\rhohat_0$, will be distinguished by the \emph{value} of $\Lik(\rhohat_0)$.  The Bayesian approach begins and ends with a distribution, which has no values.  Everything must be phrased in terms of measurable subsets (e.g., intervals), and integration over them.

Estimation algorithms transform data (observed in the past) into an estimate (which predicts the future).  In order to select the best estimate, we must logically connect the past and the future.  Frequentist methods implicitly use the frequentist axiom, while Bayesian approaches take a weighted average over all possible theories.  This averaging is particularly apropos for quantum state estimation, because density matrices have a natural convex structure.  Suppose a physicist knows that a qubit's state is $\ket{0}$ with probability $\frac13$ and $\ket{1}$ with probability $\frac23$.  He will describe it by the \emph{average} state, $\rho = \frac13(\proj{0}+2\proj{1})$ -- \emph{not} the most likely state, $\rho = \proj{1}$.

Viewed this way, the prior replaces the frequentist axiom as a connection between past and future.  This can be an advantage, for a Bayesian is capable of gracefully acknowledging that the data are \emph{not} descriptive of the true state -- that they are unlikely, or simply insufficient.  However, the price paid for this flexibility is the need to choose a prior, often without any good justification.

\section{Where do we go from here?} \label{secExtensions}

The Bayesian approach to state estimation has undeniable advantages.  It is accurate, it honestly represents the estimator's knowledge, and it conforms to quantum states' role as predictors.  \emph{Purely} frequentist approaches -- e.g., maximum likelihood as it is currently used -- cannot match these qualities.

Nonetheless, BME comes with an array of concomitant challenges.  These range from the purely practical (integration is hard) to the fundamental (how do we choose a prior?).  While some are specific to Bayesian methods, others cast doubt on the scalability of \emph{any} state estimation procedure.

\subsection{The Prior's Tale} \label{secExtensionsPrior}

Of all the problems and caveats raised by BME, none is more pressing or obvious than ``How do we choose a prior?''  BME's optimality depends on the estimator's prior matching the ``true'' distribution of unknown states.  This is fine in the rather artificial context of a state-estimating game that might be played many times, but physics experiments aren't drawn from an ensemble.  Each experiment is, as a rule, unique.

The prior is therefore a necessary fiction.  As a convenient way of representing the estimator's ignorance (either genuine, or assumed for the sake of scientific impartiality), it ought to be as \emph{uninformative} as possible.  Unitary invariance is a good first guideline (see Section \ref{secExtensionsScalability} below, however).  Over the spectrum of $\rho$, however, no \emph{uniquely} suitable measure exists.  Identifying particularly useful and non-informative priors remains an open and urgent question.

A related open question is ``What is the penalty for choosing the wrong prior?''  If accuracy is measured by an operational divergence, then BME must outperform MLE and all other methods -- \emph{if} the estimator's prior matches the distribution of unknown states.  Its robustness to a poor prior is unknown.  The optimality proof given previously is elegant in its simplicity, but precisely because of that elegance, it provides few clues to this problem.

\subsection{Practical matters} \label{secExtensionsPractical}

Every calculus student learns that integration is harder than differentiation.  Numerical integration is an active and challenging field of numerical analysis, whereas differentiation involves little more than function evalutation.  BME consists almost entirely of integration, whereas MLE is a maximization problem.  Unsurprisingly, the implementation of BME described above is roughly an order of magnitude slower than MLE.  Experimentalists, already frustrated by MLE analyses that run for a week or more \footnote{Hartmut H\"affner; private communication}, may be nonplussed.

This state of affairs may owe a great deal to the fact that MLE algorithms, unlike BME, have been developed and used for 5-10 years.  Substantial speedups are likely in the future -- precisely because numerical integration remains something of an art.  The Metropolis-Hastings algorithm already provides a tremendous advantage over na\"ive Monte Carlo, so a few more orders of magnitude may be feasible.

One reason for optimism is that the BME integral appears, in principle, to have a rather simple analytic form.  The likelihood function is a polynomial, the product of many \emph{linear} functions, of the form $\Tr(\rhohat M_i)$.  For certain priors (e.g., Hilbert-Schmidt) the resulting posterior looks a lot like a beta distribution of the form $\beta(x) = x^n(1-x)^{N-n}$.  This appears in classical estimation, and is easy to integrate.  What makes the quantum case hard is the boundary conditions.  Unlike the classical probability simplex, the quantum state-set has curved edges that are awkward to integrate over.  However, analytic solutions can be obtained for small $N$, and a general solution \emph{might} be possible.




\subsection{Scalability} \label{secExtensionsScalability}

Quantum devices exist that provide coherent control over 8 to 12 qubits \cite{HaeffnerNature05,NegrevergnePRL06}.  Twenty or thirty qubits will probably be controlled within the next five years (if only for a short time, and with limited fidelity).  The Hilbert space of a 30-qubit quantum register is enormous -- to merely store one density matrix for such a device would require just under 1 million \emph{tera}bytes of memory.  State estimation, as we know it, is impossible in this context.

Nonetheless, characterizing quantum hardware will remain important.  A quantum computer will not need state estimation; its results will appear as a computational basis state, determined by a projective measurement.  Development and testing of components, however, will depend crucially on state estimation.  It is not sufficient to know whether or not the desired state is produced; the designer will want to know the nature of the errors, so as to correct them.  Eventually, these errors need to be reduced below a fault-tolerance threshold that is probably less than $10^{-3}$.

As the states that are estimated grow larger, and the uses to which they are put become more demanding, utterly new techniques will be needed.  Unbiased estimation -- i.e., guessing the system's state without any pre-existing assumptions -- becomes exceedingly data-intensive for large Hilbert spaces.  Making use of the estimator's prior knowledge will be essential.  The Bayesian approach presented here provides a natural framework for doing so.  However, a framework for reliably representing that prior knowledge (without falling prey to self-fulfilling prophecies) will be necessary.  This reason alone would justify further study of Bayesian state estimation.

\begin{acknowledgments}
This paper is the result of more than two years of thinking about state estimation.  Much of that thinking has been done out loud, and the author is exceptionally grateful to his conversational partners.  In particular: Stephen Bartlett, Carlton Caves, Hartmut H\"affner, Patrick Hayden, Richard Gill, Daniel James, Dominik Janzing, Karan Malhotra, Serge Massar, Colin McCormick, John Preskill, Andrew Silberfarb, Rob Spekkens, and Steven Van Enk.
\end{acknowledgments}

\begin{appendix}
\section{Necessary and sufficient condition for a prior's robustness} \label{appPlausibilityTheorem}
\begin{theorem}
A prior $\pi_0(\rho)\diff\rho$ over $d\times d$ density operators is robust (and therefore guaranteed to generate full-rank estimates for any finite measurement record) if and only if its support in Hilbert-Schmidt space is \emph{not} a subset of a finite intersection of $((d-1)^2-1)$-dimensional hyperplanes that are tangent to the state-set.
\end{theorem}

\textbf{Proof: } A prior is fragile if and only if it can be annihilated by a some finite-length measurement record:  i.e., there exists $\M = \{M_1,M_2,\ldots M_N\}$ so that $\Lik(\rho) = \prod_{i=1}^N{\Tr(M_i\rho)}$ is zero on the prior's support.  $\Lik(\rho)$ is zero if and only if, for some $i$, $\Tr(M_i\rho)=0$.  
Each $M_i$ thus eliminates every $\rho$ supported on $M_i$'s null space, which has at most $(d-1)$ dimensions.  The Hermitian trace-1 matrices supported on $M_i$'s null space form a $(d-1)^2-1$ dimensional hyperplane in Hilbert-Schmidt space.  This hyperplane contains non-negative states, which are necessarily orthogonal to $M_i$, and therefore lie on the boundary of the state set.  Thus, $M_i$ eliminates all density matrices lying within a hyperplane which includes states, but does not include full-rank states -- i.e., a hyperplane that is tangent to the state-set.  The states eliminated by $\M$ are, therefore, merely the intersection of $N$ such hyperplanes, and if the prior's support does not lie within such an intersection, it cannot be eliminated.

Conversely, suppose that the prior's support \emph{does} lie within an intersection of $N$ such tangent hyperplanes.  Each hyperplane is closed under convex combination, so we can define a convex combination of \emph{every} non-negative element, $\rho_0$, which is itself an element of the hyperplane.  Since the hyperplane is tangent to the state-set, no element can lie in the interior, and so $\rho_0$ is not full-rank -- i.e., it is orthogonal to some $\proj{\psi}$.  Since $\rho_0$ is a convex combination of every state in the hyperplane, the entire hyperplane is orthogonal to $\proj{\psi}$, and is therefore eliminated by observing $\proj{\psi}$.  A measurement record consisting of the annihilating projectors for each of the $N$ hyperplanes will therefore annihilate the prior, so it is fragile. \qed

\begin{corollary} Any prior with support on a smooth curve in at least $(d-1)^2$ dimensions is robust.
\end{corollary}

\textbf{Proof: } Since the curve occupies at least $(d-1)^2$ dimensions of Hilbert-Schmidt space, it cannot be contained in a $((d-1)^2-1)$-dimensional hyperplane.  If it could be contained in a finite union of such planes, then it would not be smooth. \qed

\end{appendix}

\bibliographystyle{apsrev}
\bibliography{../bib/quantum,../bib/math,../bib/estimation,../bib/RBK}

\end{document}